\documentclass[pra,superscriptaddress,showpacs,twocolumn]{revtex4}

\usepackage{amsfonts,amssymb,amsmath}
\usepackage{graphics,graphicx,epsfig}
\usepackage{amsthm}

\def\identity{\leavevmode\hbox{\small1\kern-3.8pt\normalsize1}}

\newtheorem{theorem}{Theorem}

\newcommand{\ket}[1]{\left | #1 \right\rangle}
\newcommand{\bra}[1]{\left \langle #1 \right |}

\newcommand{\Tr}{\mathrm{Tr}}

\renewcommand{\epsilon}{\varepsilon}

\begin{document}

\title{Quantum entanglement from random measurements}

\author{Minh Cong Tran}
\affiliation{School of Physical and Mathematical Sciences, Nanyang Technological University, Singapore}

\author{Borivoje Daki\'c}
\affiliation{Faculty of Physics, University of Vienna, Boltzmanngasse 5, A-1090 Vienna, Austria}
\affiliation{Institute for Quantum Optics and Quantum Information, Austrian Academy of Sciences, Boltzmanngasse 3, A-1090 Vienna, Austria}

\author{Fran\c{c}ois Arnault}
\affiliation{Universit\'e de Limoges, 123 avenue A. Thomas, 87060 Limoges CEDEX, France}

\author{Wies{\l}aw Laskowski}
\affiliation{Institute of Theoretical Physics and Astrophysics, University of Gda\'nsk, PL-80-952 Gda\'nsk, Poland}

\author{Tomasz Paterek}
\affiliation{School of Physical and Mathematical Sciences, Nanyang Technological University, Singapore}
\affiliation{Centre for Quantum Technologies, National University of Singapore, Singapore}
\affiliation{MajuLab, CNRS-UNS-NUS-NTU International Joint Research Unit, UMI 3654, Singapore}

\begin{abstract}
We show that the expectation value of squared correlations measured along random local directions is an identifier of quantum entanglement in pure states which can be directly experimentally assessed if two copies of the state were available.
Entanglement can therefore be detected by parties who do not share a common reference frame and whose local reference frames, such as polarisers or Stern-Gerlach magnets, remain unknown. 
Furthermore, we also show that in every experimental run access to only one qubit from the macroscopic reference is sufficient to identify entanglement, violate a Bell inequality, and in fact observe all phenomena observable with macroscopic references.
Finally, we provide a state-independent entanglement witness solely in terms of random correlations and emphasise how data gathered for a single random measurement setting per party reliably detects entanglement.
This is only possible due to utilised randomness and should find practical applications in experimental confirmation of multi-photon entanglement or space experiments.
\end{abstract}


\maketitle


Quantum mechanics imposes no limits on the spatial separation between entangled particles.
This naturally leads one to ask whether observers that have never met and do not share a common reference frame can still detect effects of quantum entanglement.
One can further ask if in every experimental run each observer's local reference frame needs to be composed of a huge number of somewhat correlated elementary systems (as it is the case for Stern-Gerlach magnets, polarisers, etc.),
or if the effects of entanglement can be detected with references composed of only a few systems.

Individually both of these questions have been addressed before.
It is known that entanglement can be detected, cryptography can be realised, and Bell inequalities can be violated without a shared reference frame~\cite{PhysRevLett.104.050401,PhysRevA.83.022110,RevModPhys.79.555,PhysRevA.85.024101,SciRep.2.470,PhysRevA.86.032322,PhysRevLett.108.240501,PhysRevA.88.022327,PhysRevA.90.042336,PhysRevA.91.052118,PhysRevA.82.012304,NJP.16.043002,NJP.15.073001} and non-classical correlations can also be observed with finite-size references which are to some degree correlated~\cite{NewJPhys.11.123007,PhysRevA.79.032109,NewJPhys.13.043027}.
Here we simultaneously address both questions and 
show that observers who have independent reference frames in an unknown state can each use a single spin-$\frac{1}{2}$ of the reference per experimental run in order to detect entanglement.
If the state of the reference can be controlled a single spin-$\frac{1}{2}$ of it per experimental run will be shown to be sufficient to observe all phenomena that one can observe with macroscopic references in every experimental run.

These findings have both practical and fundamental aspects.
On the practical side, they show that entanglement detection is possible with independent reference frames
and hence observers can save on communication resources~\cite{PhysRevLett.86.4160,PhysRevA.63.052309,PhysRevLett.87.169701,PhysRevLett.87.257903} or pre-established quantum entanglement~\cite{PhysRevA.64.050302,PhysRevLett.85.2010} that would have to be consumed to correlate local reference frames.
On the fundamental side, bounded reference frames were discussed in the context of quantum-to-classical transition~\cite{NewJPhys.11.123007}, 
where it was noted that the lack of perfect reference frames leads to ``intrinsic decoherence''~\cite{NewJPhys.8.58,IntJTheorPhys.45.1189,PhysRevLett.93.240401} that might wash out all quantum features.
The present work shows that even a single qubit of a reference frame per experimental run can be used to observe Bell violation and hence reveal quantumness.


\begin{figure}[b]
\includegraphics[width=0.48\textwidth]{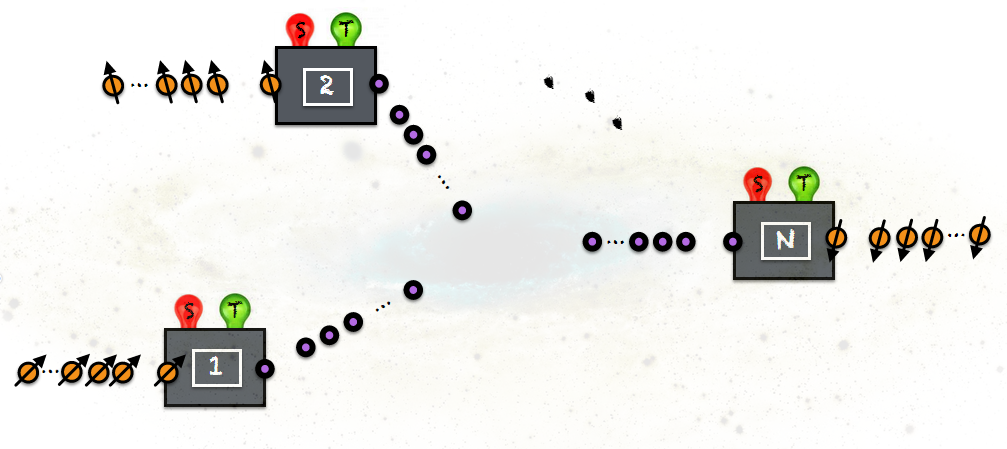}
\caption{Entanglement detection with minimal independent reference frames.
In every experimental run, each obersever, enumerated from $1$ to $N$, receives one qubit from a principal system (violet) and takes one qubit from the ensemble of identically aligned reference qubits (orange).
The reference qubits are all in the same unknown random state obtained e.g. by cooling a magnetic material below the Curie temperature.
Different observers have independent reference qubits so they can be placed even in far away arms of the galaxy.
We show that correlations between results of local total-spin measurements with outcomes denoted by $S$ and $T$ detect quantum entanglement for the principal system in an arbitrary pure state and some mixed ones.
In principle this requires infinitely many experimental runs in order to average over random directions and in order to estimate correlation functions for  fixed directions,
but we also demonstrate that this technique is useful in the presence of finite resources.}
\label{FIG_REF}
\end{figure}

\section{Experimental scenario}

Consider an experiment depicted in Fig.~\ref{FIG_REF}.
In a single experimental run the $n$th party makes use of just two qubits: one from the principal system whose entanglement is going to be estimated, and one reference qubit prepared in an \emph{unknown} pure state with Bloch vector $\vec u_n$ (measurement setting).
In order to violate a Bell inequality or detect entanglement, certain expectation values have to be estimated which require repeated measurements with the same setting.
This can be realised with the help of spontaneous magnetisation~\cite{Book.Ferromagnetism}.
Each party prepares a magnetic material that is cooled down below the Curie temperature and becomes ferromagnetic.
Spontaneous symmetry breaking causes all the spins of the material to point in the same randomly oriented direction allowing observers to use them one by one as reference qubits.
The number of spins in a magnet gives the number of experiments, $K$, with fixed settings $\vec u_n$.
For the moment we keep $K \to \infty$, and analyse the effect of finite $K$ at the end of the paper.

In each experimental run every party performs locally a total-spin measurement on the two available qubits.
The two possible outcomes correspond to the two qubits in the singlet state, $| \psi^- \rangle \langle \psi^- |$, in which case the observer assigns outcome $-3$, 
or a state in the triplet subspace, $\openone - | \psi^- \rangle \langle \psi^- |$, in which case the observer assigns outcome $+1$.
Altogether the quantum mechanical observable of the $n$th party is given by
\begin{equation}
(\openone - | \psi^- \rangle \langle \psi^- |) - 3  | \psi^- \rangle \langle \psi^- | = \sum_{j=x,y,z} \sigma_j \otimes \sigma_j,
\end{equation}
where $\sigma_j$ is the corresponding Pauli matrix.
It is now straightforward to verify that the correlation function between results $+1/\!- \!3$
obtained on the principal system in arbitrary state $\rho$ and the reference qubits reads:
\begin{equation}
E(\vec u_1, \dots, \vec u_N) = \sum_{j_1 \dots j_N = x,y,z} T_{j_1 \dots j_N}  \:.(\vec u_1)_{j_1} \dots (\vec u_N)_{j_N},
\label{E_ROT_INV}
\end{equation}
where $T_{j_1 \dots j_N} = \Tr(\rho. \sigma_{j_1} \otimes \dots \otimes \sigma_{j_N})$ are the correlation tensor elements of the state of the principal system and $(\vec u_n)_{j_n}$ is the $j_n$th component of the Bloch vector $\vec u_n$.
One recognises that Eq.~(\ref{E_ROT_INV}) is exactly the same as the correlation function between the outcomes of dichotomic $\pm 1$ observables in the presence of macroscopic reference frames in every experimental run.
Therefore, given ability to prepare (non-random) states $\vec u_n$, Eq.~(\ref{E_ROT_INV}) shows that \emph{all} quantum phenomena involving dichotomic observables on qubits and macroscopic reference frames in every experimental run can also be observed using a single qubit from a macroscopic reference per experimental run and total-spin local observables.
In particular this allows violation of an arbitrary Bell inequality with the single reference qubit per party per experimental run.
For comparison, Costa \emph{et al.} concluded that using spin coherent states as references (of the same size for every party) requires a system of dimension six in every experimental run for the violation of the CHSH inequality, and of dimension four for the Mermin inequalities, in the limit of $N \to \infty$ ~\cite{NewJPhys.11.123007}.


\section{Random correlations}

We proceed to show how correlations measured along $M$ sets of random local directions are related to quantum entanglement.
We first assume $M \to \infty$, and analyse the effect of finite $M$ at the end of the paper.

Let us represent each setting vector $\vec u_n$ in spherical coordinates $\vec u_n = (\sin\theta_n \cos \phi_n,\sin\theta_n\sin\phi_n,\cos\theta_n)$.
We define \emph{random correlations} as the expectation value of squared correlation functions averaged over uniform choices of settings for each individual observer
\begin{align}
	\mathcal{R} \equiv \frac{1}{(4\pi)^N} \int d \vec u_1 \dots \int d \vec u_N \, \,  E^2(\vec u_1, \dots, \vec u_N), \label{DEF_TINF}
\end{align}
where $d \vec u_n = \sin \theta_n d \theta_n d \phi_n$ is the usual measure on the unit sphere.
Instead of averaging over random $\vec u_n$, $\mathcal{R}$ can also be estimated from correlations along orthogonal local directions $\vec x, \vec y, \vec z$.
Let us introduce a quantity which we refer to as the \emph{length of correlations}:
\begin{equation}
\mathcal{C} \equiv \sum_{\vec u_1,\dots,\vec u_N = \vec{x},\vec{y},\vec{z}} E^2(\vec u_1,\dots,\vec u_N).
\label{DEF_C}
\end{equation}
Since $\mathcal{C}$ is invariant under local unitary operations (local rotations) \cite{PhysRevLett.93.230403}, and
the random correlations are the average of $\mathcal{C}$ over random rotations applied to local bases $\vec x, \vec y, \vec z$, it is merely a mathematical step to obtain:
\begin{align}
	\mathcal{R} = \mathcal{C} / 3^N,
	\label{REL_TC}
\end{align} 
where the factor of $1/3$ for every observer takes into account the fact that rotating one axis over a $4 \pi$ solid angle also makes the other two axes rotate over a $4 \pi$ solid angle.
The following theorem shows a universal lower bound on the random correlations in every pure state $\ket{\psi}$ of $N$ qubits.
\begin{theorem}
For all pure states of $N$ qubits, $\mathcal{R} \ge 1/3^N$.
\label{TH_CORR_BOUND}
\proof
We shall prove that $\mathcal{C} \ge 1$ for all pure states.
We begin by artificially introducing a new set of $N$ qubits prepared in the same $N$-qubit state $\ket{\psi}$.
The quantity $\mathcal{C}$ can now be linearized in the larger Hilbert space composed of initial qubits and the new qubits:
\begin{equation}
\mathcal{C} = \bra{\psi} \bra{\psi} \mathcal{S} \ket{\psi} \ket{\psi}, \label{EQN_C1}
\end{equation}
where the first ket in $\ket{\psi} \ket{\psi}$ is the state of the initial qubits $1 \dots N$, and the second ket is the state of artificially introduced qubits $1' \dots N'$.
The operator $\mathcal{S}$ acts on $2N$ qubits and is defined as:
\begin{equation}
\mathcal{S} \equiv \sum_{j_1,\dots, j_N = x,y,z} \sigma_{j_1}^{(1)} \otimes \dots \otimes \sigma_{j_N}^{(N)} \otimes \sigma_{j_1}^{(1')} \otimes \dots \otimes \sigma_{j_N}^{(N')},\label{DEF_S}
\end{equation}
where we have explicitly written the qubits on which the Pauli operators act.
In order to prove the thesis we study the eigenproblem of $\mathcal{S}$ and restrict the solutions to the subspace which is symmetric under exchange of the primed and unprimed systems.
Let us put $\mathcal{S}$ in the following form:
\begin{equation}
\mathcal{S} = H_{11'} \otimes \dots \otimes H_{NN'},
\end{equation}
where we introduced the Heisenberg hamiltonian (in the units of coupling strength)
\begin{equation}
H_{nn'} = \sum_{j_n = x,y,z} \sigma_{j_n}^{(n)} \otimes \sigma_{j_n}^{(n')},
\end{equation}
with eigenstates $\ket{00}$, $\ket{11}$, $|\psi^+ \rangle = \frac{1}{\sqrt{2}}(\ket{01} + \ket{10})$ (belonging to eigenvalue $+1$),
and $| \psi^- \rangle = \frac{1}{\sqrt{2}}(\ket{01} - \ket{10})$ (belonging to eigenvalue $-3$).
The eigenvalues of $\mathcal{S}$ are the products of these eigenvalues.
Note however that not all such products are allowed if we restrict ourselves to the symmetric subspace.
Only the eigenstates with an even number of singlet states $| \psi^- \rangle$ span the symmetric subspace.
Therefore, the allowed eigenvalues of $\mathcal{S}$ are given by
\begin{equation}
s_k = (-3)^{2k},
\end{equation}
where $2k$ gives the number of singlet pairs.
The lowest eigenvalue, $s_0 = 1$, corresponds to no singlets,
and we conclude the proof by noting that the expectation value of $\mathcal{S}$ cannot be smaller than the minimal eigenvalue.
\endproof
\end{theorem}

Our method of proof reveals that random correlations can be directly estimated from a measurement of $\mathcal{S}$ performed on two copies of the quantum state.
This is reminiscent of direct entanglement detection schemes using the two copies~\cite{PhysRevLett.98.140505,PhysRevA.75.032338,PhysRevA.78.022308,Nature.440.1022,PhysRevLett.101.260505} and suggests a deeper link between random correlations and entanglement.


\section{Random correlations and entanglement}

The length of correlations $\mathcal{C}$ and similar quantities have appeared in the literature on entanglement detection and quantification before~\cite{QuantInfComp.8.773,PhysRevA.77.062334,PhysRevA.80.042302,PhysRevLett.100.140403,PhysRevA.84.062305}.
In particular, Hassan and Joag already concluded that $\mathcal{C}$ identifies entanglement in pure states~\cite{PhysRevA.80.042302}.
However, since their derivation relies on an incorrect theorem in Ref.~\cite{QuantInfComp.8.773}, which does not seem to be easily fixable~\footnote{In Proposition 1a of Ref.~\cite{QuantInfComp.8.773} the claim at the end of the proof, that $\alpha_k$ is allowed to be zero, is incorrect because by their assumption only the correlations between \emph{all} the subsystems can be taken into account, and $\alpha_k = 0$ means that the $k$th subsystem is not measured.},
we give an alternative proof utilising our Theorem~\ref{TH_CORR_BOUND}.

\begin{theorem}
A pure state is entangled iff $\mathcal{R} > 1/3^N$.
\label{TH_ENT_ID}
\proof
Clearly, for any product state we have $\mathcal{R} = 1/3^N$ (and $\mathcal{C} = 1$).
For the converse statement assume that state $\ket{\psi}$ admits $\mathcal{C} = 1$ and decompose it in the standard basis:
\begin{equation}
\ket{\psi}=\sum_{j_1,j_2,\dots,j_N=0}^1 \alpha_{j_1j_2\dots j_N}\ket{j_1}\ket{j_2}\dots\ket{j_N}.
\end{equation}
Since $\mathcal{C}=1$, two copies of  $\ket{\psi}$, i.e. $\ket{\psi}\otimes\ket{\psi}$, lie in the subspace spanned by the tensor product of the symmetric states $\ket{00}$, $\ket{11}$, and $|\psi^+ \rangle$ only. Therefore, exchanging any qubit $i$ with its hypothetical copy $i'$ will result in the same state. For $i=1$, this leads to the relation
\begin{equation}
\alpha_{j_1j_2..j_N}\alpha_{j_{1'}j_{2'}..j_{N'}}=\alpha_{j_{1'}j_2..j_N}\alpha_{j_{1}j_{2'}..j_{N'}},
\end{equation}
for any $j_2,...,j_N,j_{2'},...,j_{N'}=0,1$. 
If we choose $j_1=0, j_{1'}=1$ and fix $j_i,j_{i'}$ for all $2\leq i \leq N$,
this relation takes the form 
\begin{equation}
\frac{\alpha_{1|J}}{\alpha_{0|J}}=\frac{\alpha_{1|J'}}{\alpha_{0|J'}}, \label{ALP_REL}
\end{equation}
where $J\equiv j_2  j_3 j_4 \dots j_N$, for any $J, J'$. 
Writing $\ket{\psi}$ in this notation we have
\begin{align}
\ket{\psi} = \sum_J \alpha_{0|J}(\ket{0}+k_J\ket{1})\otimes\ket{J},
\end{align}
where $k_J=\alpha_{1|J}/\alpha_{0|J}$ was introduced in (\ref{ALP_REL}) and shown to be independent of $J$, i.e. $k_J = k$.
Therefore the state of the first qubit is the same for every $J$ and we may rewrite $\ket{\psi}$ in a product form
\begin{align}
\ket{\psi}&=(\ket{0}+k \ket{1})\otimes  \sum_J \alpha_{0|J}\ket{J} \\
&=\ket{\Phi_1}\otimes \ket{\Phi_{2\dots N}},
\end{align}
with $\ket{\Phi_1}$ being a pure state of the first qubit and $\ket{\Phi_{2\dots N}}$ being a pure state of the last $N-1$ qubits.
Note that by construction, 
\begin{align}
1=\mathcal{C}_{\ket{\psi}}=\mathcal{C}_{\ket{\Phi_1}} \mathcal{C}_{\ket{\Phi_{2\dots N}}}=\mathcal{C}_{\ket{\Phi_{2 \dots N}}},
\end{align}
where $\mathcal{C}_{\ket{\phi}}$ is the length of correlations calculated for the state $\ket{\phi}$.
Thus we can apply induction and finally one can write $\ket{\psi}$ fully as a product state. \endproof
\end{theorem}

This theorem provides new perspectives on entanglement in pure states.
It is well known that entanglement can be verified by studying the entropy of every one-particle subsystem.
As just shown, an alternative complete characterisation exists, solely in terms of the correlation functions between \emph{all} $N$ observers (no correlations between smaller number of particles enter this characterisation).
In this sense entanglement manifests itself in \emph{full correlations}: entangled states are more correlated in random local measurements than product states.

Another characterisation of entanglement is implicit in the proofs above.
Only for product states it is possible to swap the same subsystem of the principal system and its copy without changing the whole two-copy state.
It is also worth emphasising that entanglement is detected by a two-step averaging procedure:
we need to estimate correlation functions, square them, and then average them over random measurement settings.


\section{Entanglement witness}

In principle, to determine $\mathcal{R}$, an infinite number of measurements has to be performed 
both in terms of $K$ (the resources needed to estimate correlation functions) and in terms of $M$ (the resources needed for averaging over random settings).
Recall that each party repeats $M$ times preparation of $K$ reference qubits.
We now introduce and study an entanglement witness~\cite{HoroNPT,RevModPhys.81.865} that takes the finiteness of $K$ and $M$ into account.
It can also be used to detect entanglement in some mixed states.

Let us denote by $\mathcal{R}_{M,K}$ the random correlations estimated from correlation functions measured along $M$ sets of random directions, each of which is calculated after $K$ experimental runs.
Clearly, if both $M$ and $K$ tend to infinity, $\mathcal{R}_{M,K} \to \mathcal{R}$.
We calculate the standard deviation $\Delta_{M,K}$ of the distribution of $\mathcal{R}_{M,K}$ for product states and propose the following entanglement witness:
\begin{equation}
\mathcal{R}_{M,K} > 1/3^N + 2 \Delta_{M,K} \quad \implies \quad \textrm{likely } \psi \textrm{ is ent}.
\end{equation}
Simply put, if the estimated random correlations are far away from what is expected for a product state, we most likely are dealing with an entangled state.
Compared to standard witnesses, ours is for random correlations and it can be satisfied by separable states, and yet even a single measurement setting may reveal entanglement with high confidence.
We now make this statement precise.

We calculate separately the standard deviation due to finite $M$, $\Delta_M$, and the standard deviation due to finite $K$, $\Delta_K$.
The final variance is $\Delta_{M,K}^2 = \Delta_M^2 + \Delta_K^2$.
Consider first the case of $K \to \infty$.
In Appendix~\ref{APP_RANDOM} we prove that the squared correlation of a pure product state of $N$ qubits measured along a random direction is distributed in $[0,1]$ by the density function:
\begin{align}
	\chi_N(E^2)=\frac{1}{2^N\sqrt{E^2}} \frac{(-\ln E^2)^{N-1}}{(N-1)!}.\label{DensFunc}
\end{align}
Each time a product state is measured with random settings, the squared correlation is picked from this distribution. 
After $M$ such trials, by the central limit theorem, the average of squared correlations, $\mathcal{R}_{M}$, will be normally distributed around the mean $\mathcal{R} = 1/3^N$.
The standard deviation of this normal distribution, $\Delta_{M}$, is closely related to the standard deviation $\Delta$ of the distribution (\ref{DensFunc}):
\begin{align}
	\Delta_{M} = \frac{\Delta}{\sqrt{M}} \quad \textrm{with} \quad \Delta = \sqrt{\frac{1}{5^N}-\frac{1}{9^N}}.
	\label{MINFTY}
\end{align}
For a normal distribution, there is a $95.4\%$ chance that $\mathcal{R}_{M}$ lies within $2 \Delta_{M}$ from $\mathcal{R}$.
Therefore, if the observed value of $\mathcal{R}_{M}$ is more than $2 \Delta_{M}$ away from $1/3^N$ we are $95.4\%$ sure that the state is entangled. 
This reliable state-independent entanglement witness also works for some mixed states as we show in Appendix~\ref{APP_SEP}.

To approximate the effects of finite $K$, let us denote by $E_K$ the value of the correlation function estimated from $K$ experimental runs with the same measurement settings.
For $K \to \infty$ the estimated $E_K$ tends to $E$, the quantum-mechanical prediction for a given setting.
By the central limit theorem the distribution of $E_K$ has standard deviation $\sqrt{(1-E^2)/K}$,
where $\sqrt{1-E^2}$ is the standard deviation of the binomial distribution of the product of individual measurement results.
This distribution can be used to calculate the variance of $E_K^2$ (see Appendix~\ref{APP_FINITE} for details):
\begin{equation}
\Delta_K^2 = \frac{2}{MK^2}\bigg[1- \frac{2(1-K)}{3^N}+(1-2K)\left(\frac{1}{9^N}-\Delta_M^2 \right)\bigg]
\end{equation}
where it is also assumed that $M$ is sufficiently big.
In general, in order to reveal entanglement, the number of experimental runs has to scale exponentially with the number of qubits $N$ 
as random correlations of all states are exponentially small, and their standard deviations have to be exponentially small in order to distinguish them from separable states.

Finally, we emphasise that even a single measurement setting per party suffices to confirm entanglement with a high degree of confidence.
The probability that a product state has correlation below $c = 1/3^N + \delta_N$ is given by $\int_0^{c} \chi_N(E^2) d E^2 = \Gamma(N,-\frac{1}{2} \ln(\delta_N))/(N-1)! $ with $\Gamma$ being the incomplete gamma function.
If we fix the confidence to $95.4\%$ as for the normal distribution, i.e. choose $\delta_N$ correspondingly, then we verified numerically that the probability to observe a Greenberger-Horne-Zeilinger (GHZ) correlation revealing entanglement by a single set of settings with this confidence is $26\%$ for $N=3$, and already $86\%$ for $N=10$ qubits. 
Note that this also holds for GHZ states to which local random rotations have been applied, modelling e.g. polarisation of photons propagating through fibers.
Our method also opens the door for entanglement verification in multi-photon experiments (see Appendix~\ref{APP_SEP} for exemplary detection of noisy GHZ entanglement).
For example, in the eight-photon setup of Ref.~\cite{NatPhot.6.225} a coincidence click is observed only every $6-7$ minutes.
We checked that with a single setting per party and $K = 1000$ coincidences (corresponding to about $4.5$ days of running the experiment) our method confirms entanglement with confidence $95.4\%$ ($80\%$) with probability $49\%$ (63\%).
Since this kind of entanglement detection pushes the number of measurement settings to minimum, this method is perfectly suited for entanglement detection in multipartite systems.

\section{Higher dimensions}

Although we have explicitly calculated it for qubits all our results apply to $d$-level systems.
One simply replaces in our theorems the Pauli matrices with the generators of SU($d$). 
Since there are $d^2-1$ such generators, vectors $\vec u_n$ have to be extended to $d^2-1$ dimensions as well as the corresponding sums over Pauli matrices.
By following the same lines of proofs as for qubits one finds that for all pure states of $N$ qudits
\begin{align}
(d^2-1)^N \mathcal{R} = \mathcal{C} \geq \left[\frac{d(d-1)}{2}\right]^N,
\end{align} 
and again the lower bound is achieved only by product states.
It is also straightforward to generalise these proofs to subsystems of arbitrary dimensions $d_1,d_2,\dots,d_N$.

\section{Conclusions}

We showed that pure state entanglement can be solely characterised by correlations between all involved particles and that it can be detected by measurements along random local directions.
Simply put, entangled states are more correlated than product states.
No shared reference frame is required for entanglement detection and it can even be revealed using only one qubit per experimental run from a reference in an unknown state.
Furthermore, correlations measured along \emph{one} random setting per party are shown to reveal entanglement.
This randomness empowered entanglement detection works for pure states as well as some mixed states and can be put to practical use in multiparty experiments as well as setups where frame-alignment is difficult, e.g. space experiments with photons. 
We hope that our new perspective on such a basic aspect of quantum physics as pure state entanglement will find new applications and stimulate new results in all fields that utilise it.

\acknowledgements

We thank {\v C}aslav Brukner for discussions.
This work is supported by the National Research Foundation, Ministry of Education of
Singapore Grant No. RG98/13, start-up grant of the Nanyang Technological University,
NCN Grant No. 2012/05/E/ST2/02352, and European Commission Project RAQUEL,
and Austrian Science Fund (FWF) Individual Project 2462.

\appendix

\section{Distribution of random squared correlation of product states}
\label{APP_RANDOM}

Here we show how $E^2$ of a product state of $N$ qubits is distributed if the measurement direction is chosen uniformly at random. 
We proceed by induction on the number of qubits. 
For $N=1$, without loss of generality we choose the measured state to be $\rho=\ket{0}\bra{0}$.
Arbitrary measurement is parameterised by spherical angles $(\theta, \phi)$ and expressed in terms of Pauli matrices as
\begin{align}
	\sigma(\theta,\phi)=\sin{\theta}\cos{\phi} \, \sigma_x + \sin{\theta}\sin{\phi} \, \sigma_y+\cos{\theta} \, \sigma_z.
\end{align}
The squared correlation measured along the direction $(\theta,\phi)$ is therefore
\begin{align}
	E_1^2 =  \cos^2{\theta}, \label{TsquareN1}
\end{align}
where index $1$ emphasises that only one particle is measured.
Since the measurement direction is uniformly distributed in a unit spherical shell, $\cos{\theta}$ is uniformly distributed on $[-1,1]$. 
From Eq.~(\ref{TsquareN1}) the probability density for $E_1^2\in [0,1]$ is derived to read
\begin{align}
	\chi_1(E_1^2)=\frac{1}{2\sqrt{E_1^2}}.\label{ProbDis1}
\end{align}
Now we prove that for product states of $N$ qubits squared correlation measured along uniformly random local directions, $E_N^2\in[0,1]$, is distributed according to probability density
\begin{align}
	\chi_N(E_N^2)= \frac{1}{2^N \sqrt{E_N^2}} \frac{(-\ln{E_N^2})^{N-1}}{(N-1)!}.\label{ProbDisN}
\end{align}
For $N=1$ one verifies that (\ref{ProbDisN}) returns (\ref{ProbDis1}). 
Now assume that (\ref{ProbDisN}) holds for $N=k\geq 1$. 
We shall prove that it holds for $N=k+1$ as well. 
For a product state of $k+1$ qubits the correlation factors into product of correlation for the first $k$ qubits and suitable Bloch component of the state of the last qubit:  
\begin{align}
	E_{k+1}^2 = E_k^2 \, E_1^2.
\end{align}
Since now random variable $E_{k+1}^2$ is a product of two independent random variables $E_k^2$ and $E_1^2$,
the probability density of $E_{k+1}^2$ can be calculated as~\cite{Book.Algebra}:
\begin{align}
	\chi_{k+1}(E_{k+1}^2)&=\int_{E_{k+1}^2}^{1} \chi_1(E_1^2) \chi_k \left( \frac{E_{k+1}^2}{E_1^2} \right)\frac{d E_1^2}{E_1^2}\nonumber\\
	&= \frac{1}{2^{k+1} \sqrt{E_{k+1}^2}} \frac{(-\ln{E_{k+1}^2})^{k}}{k!},
\end{align}
where the lower limit in the integral follows from $E_k^2 = E_{k+1}^2/E_1^2 \le 1$.
Thus (\ref{ProbDisN}) holds for $N=k+1$, and by induction on $N$, it holds for product state of any number of qubits. 
Using this density function it is straightforward to compute the standard deviation
\begin{align}
	\Delta=\sqrt{\langle E^4\rangle-\langle E^2\rangle^2}=\sqrt{\frac{1}{5^N}-\frac{1}{9^N}}.\label{StdProd}
\end{align}

\section{Convexity and bound on standard deviation of separable states}
\label{APP_SEP}

It is difficult to obtain standard deviation similar to (\ref{StdProd}) for general separable states. 
However, we can put an upper bound on it. 
Let $\rho$ be a separable state
\begin{align}
	\rho=\sum_i p_i \rho_i,
\end{align}
with pure product states $\rho_i$  of $N$ qubits and probabilities $p_i$. 
The correlation of $\rho$ is
\begin{align}
	E(\rho)=\sum_i p_i E(\rho_i).
\end{align}
Since $E^4$ is a convex function, $E^4(\rho)\leq \sum_i p_i E^4(\rho_i)$, we have
\begin{align}
	\Delta_\rho  &= \sqrt{\langle E^4\rangle-\langle E^2\rangle^2}\nonumber\\
	&\leq \sqrt{\sum_i p_i \langle E^4(\rho_i)\rangle} =\sqrt{\frac{1}{5^N}\sum_i p_i} = \frac{1}{5^{N/2}}.
\end{align}
Here $\Delta_\rho$ is the standard deviation of the distribution of squared random correlations of $\rho$.
Since length of correlation is a convex function, $\mathcal{C}(\rho)\leq \sum_i p_i \mathcal{C}(\rho_i)=1$.
Hence we are at least 95.4\% sure that a mixed states $\rho$ is entangled once its random correlation $\mathcal{R}_\rho$ exceeds the threshold $1/3^N$ by twice the value of $1/5^{N/2}$.
To demonstrate this idea, consider an $N$-qubit mixture of the GHZ state and white noise:
\begin{align}
	\rho = \epsilon \,\rho_{GHZ} + (1-\epsilon) \, \frac{1}{2^N}\mathbb{I}.
\end{align}
It is straightforward to verify that the random correlation of such state is given by
\begin{align}
	\mathcal{R}_\rho \simeq \epsilon^2 \frac{2^{N-1}}{3^N}.
\end{align}
Our witness reveals entanglement in $\rho$ for  $\epsilon \gtrsim \sqrt{\frac{3^N}{2^{N-2} 5^{N/2}}}$,
i.e. for exponentially small in the number of qubits admixture of the GHZ state.

\section{Standard deviation of random correlation due to finite $K$}
\label{APP_FINITE}

Here we derive the standard deviation of random correlation due to finite number of experimental runs with fixed settings, $K$, and due to finite (but large) number of random settings, $M$.
In the main text we argue that standard deviation of correlations estimated after $K$ measurement runs for fixed, say $i$th, set of settings equals $\sigma_{\epsilon_i} = \sqrt{(1 - \epsilon_i^2)/K}$,
where $\epsilon_i$ denotes the quantum mechanical correlation function for the $i$th settings.
For finite $K$, the estimated expectation value (denoted by $x$) will be normally distributed around $\epsilon_i$ within the range from $-1$ to $+1$.
We assume that $\epsilon_i$ is sufficiently small and/or $K$ is sufficiently large so that the range of the normal distribution is well within $[-1,1]$.
Then the probability density of this distribution can be written as
\begin{align}
	f(x)=\frac{1}{\sqrt{2\pi \sigma_{\epsilon_i}^2}}\text{exp}\left(-\frac{(x- \epsilon_i)^2}{2\sigma_{\epsilon_i}^2}\right).
\end{align}
The variance of squared correlations follows from the following calculations:
\begin{align}
	\langle x^2\rangle &= \int_{-\infty}^{\infty}f(x)x^2dx = \epsilon_i^2+\sigma_{\epsilon_i}^2,\\
	\langle x^4\rangle &= \int_{-\infty}^{\infty}f(x)x^4dx = \epsilon_i^4+6 \epsilon_i^2 \sigma_{\epsilon_i}^2+3\sigma_{\epsilon_i}^4, \\
	\Delta_i^2 & \equiv \langle x^4\rangle - \langle x^2\rangle^2 = 2 \sigma_{\epsilon_i}^2(\sigma_{\epsilon_i}^2+2\epsilon_i^2)\\
	&=\frac{2}{K^2}(1-\epsilon_i^2)(1-\epsilon_i^2+2K \epsilon_i^2)\\
	&=\frac{2}{K^2}\bigg[1-2(1-K)\epsilon_i^2+(1-2K)\epsilon_i^4\bigg].
\end{align}
Random correlations are additionally averaged over random measurement directions:
\begin{equation}
\mathcal{R}_{M,K} = \frac{1}{M}(\epsilon_1^2 + \dots + \epsilon_M^2). 
\end{equation}
The variance of random correlation is therefore given by:
\begin{align}
	\Delta_K^2 & = \frac{1}{M^2}\sum_{i=1}^M \Delta_i^2 \\
	& = \frac{2}{MK^2}\bigg[1- 2 (1-K)  \sum_i \frac{\epsilon_i^2}{M} + (1-2K) \sum_i \frac{\epsilon_i^4}{M} \bigg] \\
	& \approx \frac{2}{MK^2}\bigg[1-\frac{2(1-K)}{3^N}+(1-2K)(\frac{1}{9^N}-\Delta_M^2)\bigg]
\end{align}
where in the last line we assume that $M$ is large ehough so that for product states $\sum_{i=1}^M  \frac{\epsilon^2_i}{M} \to \frac{1}{3^N}$,
because it is the expectation value of squared correlations along random directions and $\sum_{i=1}^M \frac{\epsilon_i^4}{M} \to \frac{1}{9^N} + \Delta^2_{M}$, with $\Delta^2_{M}$ derived in the main text.

\bibliographystyle{apsrev4-1}
\bibliography{random_corr}

\end{document}